# The Failure of the Law of Brevity in Two New World Primates. Statistical Caveats.[1]

Short title: The Failure of the Law of Brevity in Two New World Primates.

*Ramon Ferrer-i-Cancho[2]*

*Antoni Hernández-Fernández[2,3]*

**Abstract:** *Parallels of Zipf's law of brevity, the tendency of more frequent words to be shorter, have been found in bottlenose dolphins and Formosan macaques. Although these findings suggest that behavioral repertoires are shaped by a general principle of compression, common marmosets and golden-backed uakaris do not exhibit the law. However, we argue that the law may be impossible or difficult to detect statistically in a given species if the repertoire is too small, a problem that could be affecting golden backed uakaris, and show that the law is present in a subset of the repertoire of common marmosets. We suggest that the visibility of the law will depend on the subset of the repertoire under consideration or the repertoire size.*

## 1. Introduction

World languages exhibit many statistical patterns that qualify as candidates for linguistic universals: they hold in any language where they have been tested. A popular example is Zipf's law for word frequencies, namely that the probability of the $r$-th most frequent word in a text is approximately $p(r) \sim r^{-\alpha}$, where $\alpha$ is the exponent of the law (Zipf, 1949). Languages show many other statistical regularities such as the law of abbreviation, i.e. the tendency of more frequent words to be shorter (Zipf, 1949, pp. 63; Strauss et al., 2007), a law of meaning distribution indicating that more frequent words tend to have more meanings (Zipf, 1949, pp. 28-31), or Menzerath-Altmann law stating that the longer a construct (e.g., a sentence), the shorter its components or constituents (e.g., the shorter the clauses) (Altmann 1980; Teupenhayn & Altmann, 1984; Boroda & Altmann, 1991). Statistical patterns of language defy the claim that linguistic universals are a myth (Evans & Levinson, 2009).

---

[1] This work was supported by the grant BASMATI (TIN2011-27479-C04-03) from the Spanish Ministry of Science and Innovation. We thank D. Lusseau for the opportunity to reanalyze the dolphin data (Ferrer-i-Cancho & Lusseau 2009) and S. L. Vehrencamp for making us aware of the research in ravens by Conner (1985). We are grateful to R. Dale, D. Lusseau, L. Doyle and B. Elvevåg for helpful comments.

[2] Ramon FERRER-I-CANCHO, Complexity and Quantitative Linguistics Lab. Departament de Llenguatges i Sistemes Informàtics. TALP Research Center. Universitat Politècnica de Catalunya. Campus Nord, Edifici Omega. Jordi Girona Salgado 1-3, 08034 Barcelona (Catalonia), Spain. E-mail: rferrericancho@lsi.upc.edu
[3] Antoni HERNÁNDEZ-FERNÁNDEZ, Departament de Lingüística General. Universitat de Barcelona. Gran Via de les Corts Catalanes 585, 08007 Barcelona (Catalonia), Spain. E-mail: antonio.hernandez@upc.edu

Over the last decade, many parallels between human language and the behavior of other species have been established by means of statistical 'laws' of language. Zipf's law and a parallel of the law of meaning distribution have been found in dolphin whistles (McCowan et al., 1999; Ferrer-i-Cancho & McCowan, 2009). Parallels of the law of abbreviation have been reported in chick-a-dee calls (Hailman et al., 1987), vocalizations of Formosan macaques (Semple et al., 2010) and surface behavioral patterns of dolphins (Ferrer-i-Cancho & Lusseau, 2009). Beyond the realm of animal behavior, qualitative agreement with Menzerath-Altmann law has been found in genomes at different levels of organization (Ferrer-i-Cancho & Forns, 2009; Hernández-Fernández et al., 2011; Li, 2012). The focus of the present article is shedding light on recent research on the law of brevity or abbreviation in two New World primates (Bezerra et al., 2011).

Zipf's law of brevity, i.e. the tendency of more frequent words to be shorter (Zipf, 1936; Zipf, 1949; Straus et al., 2007), can be generalized as the tendency of more frequent elements to be shorter or smaller, following an abstraction akin to the one that led to the current formulation of Menzerath-Altmann law in terms of constructs and constituents (Altmann, 1980). Compression (Salomon, 2007; Dawkins, 1976), the information theoretic principle of assigning shorter codes to more frequent elements, is argued to underlie the generalized law of brevity across species (Zipf, 1949; Ferrer-i-Cancho & Lusseau, 2009), and is independent from modality or whether the behavior is communicative or not. The law is not a perfect rule (Zipf, 1949; Strauss et al., 2007; Ferrer-i-Cancho & Lusseau, 2009; Semple et al., 2010) and its presence does not imply that compression is the only principle involved (Bezerra et al., 2011).

Recently, it has been argued that the failure to find the law in the vocalizations of two New World primates: common marmosets and golden-backed uakaris, means that the law is not widely applicable in animal communication (Bezerra et al., 2011). In contrast, we propose here that the law of brevity has not surfaced in these two primates due to two not necessarily exclusive reasons: insufficient sampling and mixing of call types with distinct compression pressures.

The organization of the remainder of the article is as follows. Section 2 presents a summary of the presentence of the law of brevity in humans and other species unifying methodologies. Section 3 unravels some limits of the methods used to detect the presence of the law in a given species. Section 4 reviews important statistical issues for research on the law of brevity across species.

## 2. The Law of Brevity in Various Species

Here, we aim to provide a methodologically homogenous summary of the statistical evidence about the law of brevity in various species that will be the basis of a meta-analysis in the next section.

### 2.1. Materials and Methods

### 2.1.1. Materials

The frequency and duration data for common marmosets and golden-backed uakaris that is used in this article comes from the electronic supplementary material of Bezerra et al. (2011), which, in both cases, covers a subset of the whole repertoire of those species (see Table 1). The frequency and duration data for common ravens comes from the main text of Conner (1985), who did not study the law of brevity in his article. Unfortunately, the full data from the pioneering research on the law of brevity in chick-a-dee calls (Hailman et al., 1987; Hailman et al., 1985; Ficken et al., 1978) is not available for the kind of reanalysis intended here.

In humans, the law of brevity has been typically studied in written language (e.g., Zipf, 1949; Strauss et al., 2007). The use of written language could give a prior advantage to human language in terms of the degree of adherence to the law of brevity. In order to study the law of brevity in oral human language and to allow for a fairer comparison with animal data, all available data (15 April, 2011) in the Childes Database (MacWhinney, 2000) for the following languages was used: Russian, Croatian, Greek, Swedish and Indonesian. A selection of corpus of American English (Bloom, Bates, Brown, HSLLD and MacWhinney) and Spanish (Aguirre, BecaCESNo, DiezItza, Irene, OreaPine, Ornat and SerraSole) had to be used to keep the total number of words close to the other languages examined. All the corpora are freely available at http://childes.psy.cmu.edu.

For the analysis of each language, all the speakers were considered (children and adults, regardless of their role). This condition was used to approximate the mixing of individuals of the majority of animal studies reviewed here. For simplicity, the units of the analysis were word forms.

**Table 1.** Summary of the results of the exploration of the law of brevity in various species. $r_s$ and $p$ were rounded to leave, respectively, two and one significant digits.

| Species | $n$ | Law of brevity | $r_s$ | $p$ | Reference |
|---|---|---|---|---|---|
| Golden-backed uakaris | 7 (9) | No | -0.36 | 0.4 | Bezerra et al. (2011) |
| Common marmosets | 12 (17) | No | 0.056 | 0.9 | Bezerra et al. (2011) |
| Common ravens | 18 | No | -0.060 | 0.8 | This article. |
| Dolphins | 31 | Yes | -0.51 | 0.003 | Ferrer-i-Cancho & Lusseau (2009) |
| Formosan macaques | 35 | Yes | -0.43 | 0.01 | Semple et al. (2010) |
| Humans | >> 35 | Yes | (-0.26, -0.17) approx. | << 0.001 | Table 2 of this article. See also: Zipf (1935), Zipf (1949), Strauss et al. (2007). |

**Table 2.** The correlation between frequency and length in oral human language.

|            | Tokens    | Types  | $r_s$  | $p$          |
|------------|-----------|--------|--------|--------------|
| Russian    | 54,104    | 7,908  | -0.204 | $< 10^{-9}$  |
| Spanish    | 980,797   | 27,479 | -0.229 | $< 10^{-9}$  |
| Croatian   | 298,038   | 15,381 | -0.269 | $< 10^{-9}$  |
| Greek      | 52,158    | 4,203  | -0.191 | $< 10^{-9}$  |
| Swedish    | 511,191   | 19,164 | -0.161 | $< 10^{-9}$  |
| US English | 2,674,937 | 24,101 | -0.227 | $< 10^{-9}$  |
| Indonesian | 2,417,587 | 30,461 | -0.227 | $< 10^{-9}$  |

## 2.1.2. Methods

The results of the correlation between frequency and length for dolphins in Table 1 were obtained by a reanalysis of the data of Ferrer-i-Cancho & Lusseau (2009), where a Pearson correlation test was used, through a Spearman rank correlation test.

Notice that the Spearman rank correlation test is a very powerful tool to study and compare the law of brevity from heterogeneous sources. For instance, Semple et al. (2010) studied the law of brevity as a relationship between mean call type duration and frequency while in our analysis of human data (Table 2) we are studying the relationship between word length in letters and frequency. The results of the Spearman rank correlation test on human data will not change if it is assumed that mean word duration is a strictly monotonically increasing function of duration. In the end, the relevant unit of measurement under a compression or coding efficiency hypothesis is not duration or length but the energetic cost of the word. The results of the Spearman rank correlation tests will not change if that cost is a strictly monotonically increasing function of duration in seconds or length in discrete units.

## 2.2. Results

Table 1 summarizes the results of the study of the law of brevity in various species using the same notation as in Bezerra et al. (2011): $n$ for the repertoire size considered in the references of the last column, $r_s$ for the Spearman rank correlation between frequency and duration/length and $p$ for the $p$-value. In parenthesis, the size of the whole repertoire according to Bezerra et al. (2011) is also shown. Table 2 summarizes the results of the study of the law for word frequencies and their length (in letters) in various languages.

## 3. The Pressure for Brevity Can Be Hidden

A key issue in the quest for exceptionless universals (Evans & Levinson, 2009) or universal principles (Köhler, 2005; Zipf, 1949) is understanding the limits and other subtleties of the statistical methods that are being used.

## 3.1. Pressure for Brevity without Statistical Significance

Table 1 shows that the law of brevity has not been found in two New World primates (golden-backed uakaris and common marmosets) and common ravens. The law has therefore not been found in 3 out of 6 cases thus far. The small repertoire sizes of these three species may have caused type II statistical errors when trying to reject the null hypothesis that frequency and duration are unrelated. We consider the probability that these three cases coincide with the three cases that have the smallest repertoire, as it is the case according to Table 1, simply by chance. This probability is $q = (3!3!)/6! = 0.05$. Thus the null hypothesis that the coincidence is accidental can be rejected at a significance level of 0.05.

Notice that our meta-analysis for type II errors is conservative: we consider that all the languages where the law of brevity has been reported (e.g., Zipf, 1935; Strauss et al., 2007) have collapsed into the category "humans" in Table 1. If we considered that each of the $L$ languages where the law has been found must contribute separately, the *p*-value would become $q = (3!(2 + L)!)/(5 + L)!$, which cannot exceed the *p*-value of our initial calculation, i.e. 0.05, for $L \geq 1$. The fact that $q = 6/((5 + L)(4 + L)(3 + L))$ indicates that $q$ drops as $L$ grows. For simplicity, let us consider only the positive reports of the law in Table 2, which gives $L = 7$. Then we would have $q = 6/(12 \cdot 11 \cdot 10) \approx 0.0045$, namely a stronger support for the hypothesis of type II errors that could, nevertheless, be unfairly biased by human languages.

Table 3 shows different values of $n^*$, the minimum repertoire size that is needed to achieve significance in correlation tests for different significance levels $\alpha$ (see Appendix for the precise mathematical argument) depending on whether the test is one-tailed ($k = 1$) or two-tailed ($k = 2$). Absence of ties was assumed for calculating Table 3. Knowing that the law of brevity is not a perfect rule, we suggest that the repertoire size of golden-backed uakaris, 7 calls (Bezerra et al.,(2011)), is dangerously close to 5, the value of $n^*$ for a two-tailed correlation test at a significance level of 0.05 (Table 3). There are no ties in the datasets of the two New World primates that we reanalyze from Bezerra et al. (2011).

**Table 3.** The minimum sample size for significance ($n^*$) versus the significance level ($\alpha$).

| | $n^*$ | |
|---|---|---|
| $\alpha$ | $k = 1$ | $k = 2$ |
| 0.05 | 4 | 5 |
| 0.01 | 5 | 6 |
| 0.001 | 7 | 7 |
| 0.0001 | 8 | 8 |

## 3.2 The Law of Brevity Hidden in a Subset of the Repertoire

The fact that the law of brevity has not been found for the whole repertoire does not a priori imply that it cannot be found for a subset of the repertoire. Simpson's paradox indicates that statistically significant correlations may emerge when the sample is partitioned according to a certain criterion (DeGroot, 1989).

Notice that Semple et al. (2010) and Bezerra et al. (2011) study the correlation between the frequency of occurrence $f$ of a call type and its mean duration $<d> = D/f$, where $D$ is the sum of the durations of the occurrences. Since $<d> = D/f$, it is mathematically convenient to consider the dependency between $D$ and $f$ (Hernández-Fernández et al., 2011; Li, 2012), which is shown in Fig. 1 for common marmosets, to investigate the possibility of different compression pressures. Error bars indicate standard errors. The standard error of $D$ was inferred from the standard error of $<d> = D/f$ (that is available in the supplementary online information of Bezerra et al. (2011)) through $\sigma(D) = f \sigma(<d>)$. For this species, Fig. 1 A suggests two different clusters of call types: a low $D$ cluster ending at the 6-th call type with the smallest $D$ and a high $D$ cluster beginning at the 7-th call type with the smallest $D$. The cluster boundaries can be determined quantitatively. If $D_i$ is defined as the $i$-th call type with the smallest total duration, the boundary between clusters is defined by the value of $i$ that maximizes the difference in order of magnitude between consecutive total durations, i.e. $\Delta_i = \log(D_i/D_{i-1})$, where log is a natural logarithm. Table 4 shows that $\Delta_7$ is maximum. $\Delta_i$ was rounded to leave only two decimal digits.

A negative correlation between $f$ and $<d>$ is found for the low $D$ cluster ($n = 6$, $r_s = -0.886$, $p = 0.019$) but not for the high $D$ cluster ($n = 6$, $r_s = -0.086$, $p = 0.872$). The low $D$ cluster contains the call types "tisik", "very brief whistle", "chatter", "submissive squeal", "egg" and "tse". Unfortunately, the same analysis cannot be extended to golden-backed uakaris because the repertoire studied considered by Bezerra et al. (2011) is too small. The possibility that the law of brevity has emerged in the low $D$ cluster for a trivial reason will be examined further.

**Table 4.** The difference in order of magnitude between the $i$-th and the $(i - 1)$-th signal with the smallest $D$.

| $i$ | $\Delta_i$ | $i$-th signal with the smallest $D$ |
|-----|------------|-------------------------------------|
| 1   | -          | Tisik                               |
| 2   | 0.25       | Very brief whistle                  |
| 3   | 0.25       | Chatter                             |
| 4   | 0.06       | Submissive squeal                   |
| 5   | 1.03       | Egg                                 |
| 6   | 0.08       | Tse                                 |
| 7   | 1.35       | Brief phee call level 3             |
| 8   | 0.18       | Brief phee call level 2             |
| 9   | 0.47       | Brief phee call level 1             |
| 10  | 1.07       | Long phee call                      |
| 11  | 0.34       | Twitter                             |
| 12  | 0.50       | Trill                               |

A negative correlation between frequency and duration could be obtained simply when the expectation of $D$ given $f$ is constant, e.g., $D$ and $f$ are independent (Hernández-Fernández et al., 2011; Li, 2012), giving $<d> = a/f$, where $a$ is a constant, which means a very strong correlation between $<d>$ and $f$. A constant expectation for $D$ given $f$ has been excluded as an explanation of the law of brevity in the vocalizations of Formosan macaques by showing that $D$ and $f$ are indeed correlated (Semple et al., 2012). The problem of constant expectation does not concern the presence of the law in dolphin surface behavioral patterns (Ferrer-i-Cancho & Lusseau, 2009) or human word lengths (Table 2) because in those cases the law is studied between fixed length $\lambda$ (in letters for human words) and frequency and then $D = \lambda f$, i.e. $D$ and $f$ are of course, correlated. Concerning Formosan macaques, a dependency between $D$ and $f$ in the low $D$ cluster could not be supported by a two-sided Spearman rank correlation test ($n = 6$, $\rho = 0.6$, *p*-value = 0.2) but could be sustained by a two-sided Pearson rank correlation test ($n = 6$, $r = 0.97$, *p*-value = 0.04) at a significance level of 0.05.

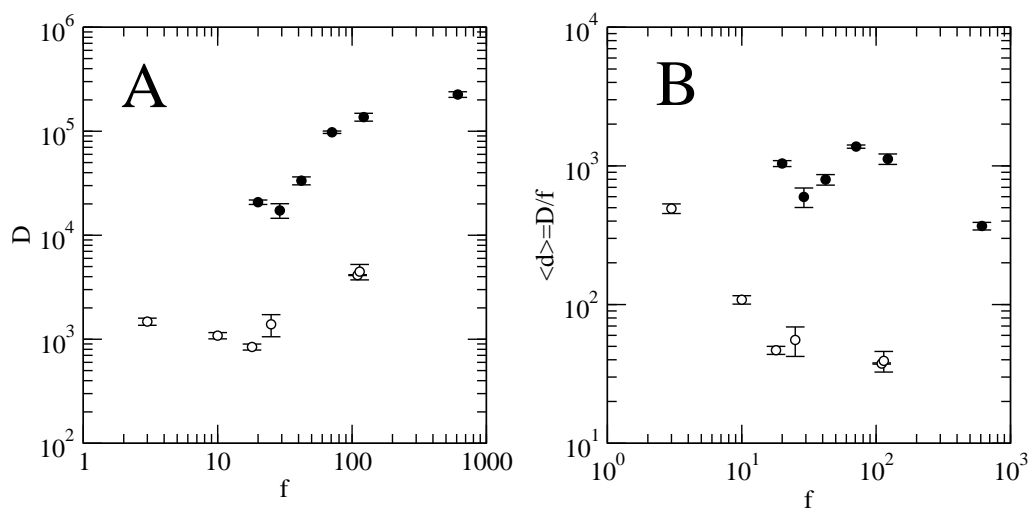

The ability of the Pearson correlation statistic to unravel a significant correlation between *D* and *f* in the low *D* cluster may suggest that the dependency between both is trivially linear, i.e. $D = af + b$, where *a* is the slope and *b* is the intercept. However, such a linear dependency is unlikely. To see it, notice first that, by the definition of total duration *D*, $b = 0$ as *D* must be zero when $f = 0$. Second, if $b = 0$ then $\langle d \rangle = D/f = a$ but a significant correlation between $\langle d \rangle$ and *f* has already been shown.

## 4. Discussion

Concerning the failure to find the law of brevity in two New World primates, it was said that "brevity is likely to be vital in memorizing the components of a large vocabulary" as in the case of human languages (Bezerra et al., 2011). However, if a very large repertoire size was crucial for an agreement with the law, one would expect not to find the law of brevity in the small repertoires of macaques and dolphins (Semple et al., 2010; Ferrer-i-Cancho & Lusseau, 2009), which are various orders of magnitude smaller than those of human languages (Table 1) or even within a very small subset of the repertoire of common marmosets. Coding efficiency is important even in small repertoires.

The supported hypothesis of false negatives (type II errors) for the law of brevity has various implications for animal behavior research. It should be checked if the vocalization types that were excluded from the analysis of the two New World primates (see Table 1), 5 out of 17 in marmosets and 2 out of 9 in uakaris (Bezerra et al., 2011) would change the results on the law of brevity. Besides, Table 3 shows that the minimum repertoire size needed for significance, $n*$, is smaller for one-tailed tests. Indeed, to test the hypothesis test of the law of brevity we only need to show that the correlation between frequency and duration is significantly small. Thus, we propose that one-tailed tests are adopted for the study of the law of brevity in other species especially for small repertoires.

Our meta-analyses suggest that the law of brevity and the underlying principle of compression may not be detected by conventional statistical approaches in a particular species if the repertoire is not large enough. Although type II errors are normally overcome by increasing the sample size, the invisibility of the law in a certain species may be intrinsic, or hard to avoid, because further sampling can improve the accuracy of the correlation but not the size of the repertoire, which is normally fixed *a priori*. We suggest that the visibility of the law in a given species will vary according to the size of the repertoire or according to our ability to identify the subsets of the repertoire where the pressure for brevity is high enough. At present, there is insufficient evidence for golden-backed uakaris and common marmosets as true exceptions to a statistical universal that is known as the law of abbreviation (Zipf, 1949).

In this article we have approached the problem of the law of brevity from a quantitative linguistics point of view. We have reviewed various statistical issues that are important for future research from a biological or ethological perspective. Concerning marmosets, it has not escaped our attention that all the possible long distance calls in the dataset that we reanalyzed ("long phee call" and "twitter" (Bezerra, 2011)) belong to the high $D$ cluster, the cluster where the law of brevity is not found. Pressure for brevity is reduced for long-distance communication (Slabbekoorn, 2006). Information theory indicates that redundancy and increasing duration, in particular, facilitate transmission (Cover & Thomas, 2006, pp. 184) and enhancing transmissibility is more important for long-range than for short range communication (Wiley, 2009, pp. 829; Slabbekoorn, 2006). The elongation of signals (the opposite of abbreviation) is known to be a solution adopted by common marmosets to fight against background noise (Brumm et al., 2004). However, determining the bioacustical or biological factors hiding the law of brevity in golden-backed uakaris or leading to the two clusters that we have discovered using statistical arguments should be the subject of a further research.

**APPENDIX**

We investigate the rank correlation test as a particular case of randomization test (Sokal & Rohlf, 1995, pp. 803-819) to shed light on the limitations of rank correlation tests with small repertoires. We start with the calculation of the lower bonds for the $p$-value $p$ of one-tailed and two-tailed rank correlation tests. Imagine that one wants to calculate $\rho(X, Y)$, the correlation between two vectors $X = x_1,...,x_i,...,x_n$ and $Y = y_1,...,y_i,...,y_n$, and that there are no ties either in $X$ or in $Y$. $X'$ is used to refer to a permutation of $X$. In a one-tailed rank correlation test, $p$ is the proportion of permutations of where $\rho(X', Y) \leq \rho(X, Y)$ (or $\rho(X', Y) \geq \rho(X, Y)$ depending on the tail of interest). We have $p \geq 1/n!$, where $n$ is the repertoire size and 1 is the minimum number of permutations of $X$ yielding a correlation at least as large (or at most as low, depending on the tail of interest) as that between $X$ and $Y$. The permutation $X' = X$ is at least one of those permutations.

In a two-tailed rank correlation test, we have $p \geq 2/n!$, where $n$ is the repertoire size and 2 is the minimum number of permutations yielding $X'$ such that $|\rho(X', Y)| \geq |\rho(X, Y)|$. The factor 2 comes from the fact that there are at least two of these permutations: $X'=X$ and $X'=X''$, where $X''$ is the inverse of $X$, i.e. $x''_i = x_{n-i+1}$. Notice that $\rho(X, Y) = - \rho(X'', Y)$ and thus $|\rho(X, Y)| = |\rho(X'', Y)|$.

For statistical significance, it is required $p \leq \alpha$, where $\alpha$ is the significance level. $n*$ is defined as the minimum value of $n$ needed to achieve significance. Table 3 shows the different values of $n*$ as a function of different values of $\alpha$ for one-tailed and two-tailed tests. $n*$ is the minimum value of $n$ that satisfies $k/n! \leq \alpha$, where $k$ is the number of tails ($k = 1$ for one-tailed and $k = 2$ for two-tailed).